\documentclass[preprint,12pt]{elsarticle}
\usepackage{soul}
\usepackage{xspace} 
\usepackage{color, colortbl}
\usepackage[table,xcdraw]{xcolor}
\usepackage{enumerate}
\usepackage{array}
\usepackage{booktabs}
\usepackage{multirow}
\usepackage{amsmath}
\usepackage{cleveref}
\usepackage{tikz, times}
\usepackage{ifthen}
\usepackage{amssymb}
\usetikzlibrary{mindmap, backgrounds, trees}

\definecolor{Gray}{gray}{0.9}
\newcommand{\E}{\textsc{Company}\xspace}
\newcommand{\BTH}{\textsc{BTH}\xspace}

\newcounter{takeawaycnt}
\newcommand{\takeaway}[1]{ %
    \stepcounter{takeawaycnt} %
   \begin{center}
    \begin{tikzpicture}
        \node[anchor=text,text width=\columnwidth-1.2cm, draw, rounded corners, line width=1pt, fill=black!5, inner sep=5mm] (big) {\\#1};
        \node[draw, rounded corners, line width=.5pt, fill=white, anchor=west, xshift=5mm] (small) at (big.north west) {\textbf{Takeaway \arabic{takeawaycnt}}};
    \end{tikzpicture}
   \end{center}
}
\newboolean{final}
\setboolean{final}{true} 
\newcommand{\review}[1]{\ifthenelse{\boolean{final}}{#1}{\textcolor{blue}{#1}}}

\journal{Journal of Systems and Software}

\begin{document}
\usetikzlibrary{positioning}
\begin{frontmatter}



\title{When Traceability Goes Awry: an Industrial Experience Report}


\author[inst1]{Davide Fucci}

\author[inst1]{Emil Al\'egroth}
\author[inst2]{Thomas Axelsson}

\affiliation[inst1]{organization={SERL, Blekinge Institute of Technology},
            addressline={Valhallavagen 1}, 
            city={Karlskrona},
            postcode={37141}, 
            country={Sweden}}
\affiliation[inst2]{organization={COMPANY},
            addressline={Address}, 
            city={City},
            postcode={Postcode}, 
            country={Country}}
\begin{abstract}
The concept of traceability between artifacts is considered an enabler for software project success.
This concept has received plenty of attention from the research community and is by many perceived to always be available in an industrial setting.

In this industry-academia collaborative project, a team of researchers, supported by testing practitioners from a large telecommunication company, sought to investigate the partner company's issues related to software quality.
However, it was soon identified that the fundamental traceability links between requirements and test cases were missing.
This lack of traceability impeded the implementation of a solution to help the company deal with its quality issues.

In this experience report, we discuss lessons learned about the practical value of creating and maintaining traceability links in complex industrial settings and provide a cautionary tale for researchers. 
\end{abstract}


\begin{highlights}
\item We report the design of an empirical study aimed at identifying and addressing the issue of quality inflation.  
\item We reflect on how the lack of simple traceability links between requirements and test cases hindered the development of the proposed solution.
\item Based on our experience, we present takeaways and considerations for researchers and practitioners about the role and value of traceability in industrial settings.  
\end{highlights}


\end{frontmatter}


\section{Introduction}
\label{sec:introduction}


The concept of traceability refers to links between different software artifacts, such as requirements, source code, and test cases.
Establishing traceability within requirements helps the organization manage dependencies. In contrast, traceability between requirements and test cases, also referred to as alignment~\cite{unterkalmsteiner2014taxonomy}, is necessary to measure coverage and ensure that the product fulfills customers' needs with a given degree of quality. 
Moreover, the traceability between source code and test cases enables the analysis of the impact when changes occur.

In this study, Blekinge Institute of Technology (\BTH) collaborated with a large telecommunication company (hereafter, \E\footnote{We omitted the \E name due to their wish to stay anonymous.}), with the original intent to study \E's software quality.
Specifically, the researchers collaborated with the organization within \E in charge of the verification before release (i.e., TestOrg).
The initial aim of this collaboration was to investigate the challenges related to the state-of-practice of the \E quality assurance process.
TestOrg was a suitable partner in this project since they are responsible for the quality assurance of all \E products and can be considered the first user of the product once it leaves the development phase.
\BTH and TestOrg organized a series of workshops to elicit pain points regarding software quality.
One outcome of such workshops was the perceived mismatch between the organization level of quality assurance activities and the actual observed quality. We termed this difference \textit{quality inflation}.
The study of the quality inflation phenomenon at \E became the refined objective of our industrial collaboration.

The researchers analyzed the artifacts TestOrg uses to perform their activities, such as requirement specifications, issue reports, and test cases.
To observe the symptoms and connect them to possible root causes of quality inflation, the researchers needed to connect the requirements to other artifacts in the development process. 
Therefore, the existing traceability links and the possibility of establishing new ones play a significant role for detecting quality inflation.
However, we found that such traces were generally lacking and that reverse engineering these traces from existing artifacts was not only time-consuming but also, in most cases, not possible.

In this paper, we report our experience in detecting quality inflation at \E and how the investigation of this phenomenon failed due to the lack of traceability between artifacts.
This experience serves as a case for academia to consider for future research and provide an example for practitioners of the challenges that can arise once traceability between software development artifacts is lacking. 

The rest of the paper is organized as follows: \Cref{sec:background} presents an overview of the existing literature on traceability in software projects and its application to different artifacts in the software development lifecycle. 
\Cref{sec:settings} presents the industrial context of our study, and \Cref{sec:traceability} shows our experience in trying to establish traceability links in a large-scale organization to investigate a particular aspect of software quality---i.e., quality inflation. 
\Cref{sec:discussion} reports a discussion of our experience and takeaways for researchers and practitioners.
Finally, \Cref{sec:conclusion} concludes the paper. 

\section{Software artifacts Traceability}
\label{sec:background}

The body of research in traceability is vast, with early works published in the 70s---e.g., Randell~\cite{randell1968towards}.
Since then, the concept has been explored in both industry and academia through empirical studies and summarized in systematic literature reviews~\cite{CAK21,TWL21,NDS13}.
Traceability is essential to govern the software development process, to manage complexity, and mitigate costs~\cite{watkins1994and,kukkanen2009applying}.
However, research has also established that maintaining traceability links of high quality over time can be a costly process~\cite{cleland2003event}.
Therefore, research has started focusing on traceability automation~\cite{hayes2007requirements}.

Among the secondary studies, Mustafa and Labiche performed a literature review that identified a lack of tools for tracing heterogenous artifacts~\cite{mustafa2017need}.
Conversely, a review by Tufail et al.~\cite{tufail2017systematic},  identified seven models, ten challenges, and 14 tools for traceability.
The review by Javed and Zdun examined the connections between traceability and software architecture~\cite{javed2014systematic}. In contrast, Santiago et al. looked at managing traceability in the context of model-driven engineering and the associated complexities~\cite{santiago2012model}.
According to a recent mapping study surveying 63 papers between 2000 and 2020, traceability involving testing artifacts and related activities is the least investigated~\cite{TWL21}.
Moreover, according to the authors, few tools support traceability in software testing activity~\cite{TWL21}.
\review{This conclusion is drawn, despite the numerous research on the tools that either explicitly focus on, or support, traceability in software development.
Examples of such tools include PLM/ALM which, according to a study by Ebert~\cite{ebert2013improving}, was beneficial to define traceability for testing purposes.
The project management tool \texttt{DOORS} is commonly used in the industry and is used as a driver for research into, for example, automated traceability links~\cite{lin2006poirot} and rich traceability~\cite{dick2002rich}.
Another example is \texttt{Enterprise architect}---a large-scale modelling framework that can be used to model traceability between several aspects of development, including assets, processes, and the organization~\cite{tang2007rationale}.
Although used successfully in practice, these tools share a human component.
Thereby, their success is tied to how rigorously they are used.
}

Tian et al.~\cite{TWL21} show that traceability in different software development activities has rarely been evaluated in industrial settings (i.e., 16\% of the reviewed primary studies).
However, despite several studies commending traceability as a prerequisite for a successful software project (and conversely pointing to a lack of traceability as a factor leading to failure~\cite{FWK17}), there is no empirical evidence, to the best of our knowledge, that supports these claims in the industrial context.

Traceability is often discussed as a sequential trace from requirements to code, and from code to test cases.
Another critical dimension is alignment---i.e., the traceability between requirements and tests.
Unterkalmsteiner et al.~\cite{unterkalmsteiner2014taxonomy} proposes a taxonomy for requirements and test alignment (REST).
They show a method for designing contextual taxonomies of alignment, including concepts for how to reason about the establishment of traceability links from requirements specifications to design, development, and testing.

Another concept related to traceability is change management~\cite{borg2017software}---i.e., the idea that when a software artifact is changed, all associated artifacts need to be identified and updated accordingly.
The responsibility for this task is often delegated to the development team, and its complexity is affected by the already available traceability links.
In their case study, Borg et al. found that developers prefer flexible forms of information rather than formal information, including traceability information, when dealing with changes~\cite{borg2017software}.

\review{Both the research on requirements-tests alignment and change management demonstrate an academic idea of the \textit{importance} of traceability in software development.
This suggestion, although reasonable, does not account for the costs of keeping traceability and alignment up-to-date in complex industrial settings.}

\section{Investigation of Software Quality within \E}
\label{sec:settings}
\BTH and \E are collaborating on a research project in the area of software quality, with a focus on automated software testing.
This project is enabled by \BTH's approach to Industry-Academia collaboration and technology transfer based on the model proposed by Gorscheck et al.~\cite{GGL06}.
Since the project is performed in co-production, \BTH began by identifying the demands and constraints imposed on TestOrg by the process used to develop and deliver software at \E.
TestOrg is in charge of verifying digital business solutions (DBS)---e.g., online charging systems, mobile financial services, and service catalog manager---for telecommunication operators worldwide.

\subsection{Preliminary Workshops and Research Objectives}
\label{sec:objectives}
In the Fall of 2019, \BTH organized two explorative workshops at \E to understand the organizational challenges in the area of testing and quality assurance related to automation.
The first workshop (Workshop 1) involved members of the quality assurance  (QA) team and product managers, whereas the second (Workshop 2) involved developers, testers, and operations personnel.
We assumed that the challenges highlighted within the two groups would differ based on norms and values~\cite{lenberg2016initial,lenberg2018psychological}.

The workshops were organized into three parts.
First, the participants answered the question \textit{``What are the challenges your team and organization are currently facing with test automation?''} in brief statements on post-it notes (several answers are possible) in a time-boxed exercise (15 minutes). 
In the second part, each participant pitches their answers to the entire group and engages in a discussion.
As the participants read and connect their answers, the researchers cluster the notes according to themes emerging during the discussion.
This part of the workshop is not time-boxed to give everyone the possibility to voice their ideas.
In the third part, the researchers present the themes that emerged during the discussions (i.e., clusters) and, in real time, validate with the participants the correct interpretation of their answers.   
We identified eight themes associated with QA and Test Automation challenges in Workshop 1 and six in Workshop 2. 

The researchers prioritized the themes that emerged during the workshop in collaboration with the TestOrg leaders and test managers, taking the results from the developers as supporting input.
The rationale for this decision was that the emerging challenges on the management level were of more general and larger complexity; thereby, they incorporated several of the challenges expressed by the developers.
\BTH presented the results to the TestOrg team, and during a discussion structured around the identified challenges, TestOrg provided further input to prioritize them.

Finally, \BTH and TestOrg set out to study \textit{quality inflation}---a mismatch between the \review{perceived effort} in quality assurance activities performed by TestOrg and the quality observed in use.
\review{We use the term inflation to indicate that the effort in quality assurance activities is artificially increased; hence, creating such mismatch.}
TestOrg perceived that the motivations for quality inflation were related to both technical and human factors. 
For example, during Workshop 1, participants pointed out the misalignment between the metrics used by management and the ones used by developers to evaluate quality. 
TestOrg managers felt that the team would intentionally write ``happy'' test cases and defer more thorough testing activities down the development pipeline. 
During Workshop 2, developers felt that quality goals are not explicit and communicated with poor rationale.

Focusing on the technical causes first, \BTH researchers collaborating with TestOrg focused on the following objectives.
\begin{description}
\item[O1.] Evaluate the quality of the test suites associated with different types of Business Requirements (BR). A BR is a high-level requirement for the product, which is later broken down into smaller requirements. During Workshop 1, it emerged that some BRs within the DBS system are perceived to work well since they have few issue reports (IR) associated with them.
We aim to locate such BRs, characterize them based on aspects such as their history and the history of the associated IR and test cases, and finally compare them to those BRs that are considered \textit{troublesome} by the development and QA teams.
This analysis aimed to show TestOrg possible root causes for quality inflation. 
\item[O2.] Correlate the quality of the automated test cases with the quality of the test specification (e.g., test smells---\review{sub-optimal design of test code~\cite{VMV01}}). This objective assumes that better-specified test scenarios lead to better test cases.
\item[O3.] Evaluate the effects of the triaging process on perceived quality. During Workshop 1, it appeared that the number of open IRs plays a crucial role in establishing the managers' perception of quality (e.g., when a BR receives many IRs, or IRs for a BR are often reopened, that BR may be perceived as low-quality).
However, IRs go through a triaging process to decide if (and to what extent) they will be tested (either with additional tests or during regression testing).
We wanted to study such triaging process as it directly impacts the number of open/closed IRs and, in turn, the perceived quality of a BR.
\item[O4.] Evaluate quality with respect to functional vs. non-functional BR features, such as performance. During Workshop 1, participants agreed that an area in which TestOrg should improve is testing of non-functional requirements. On the other hand, in Workshop 2, developers pointed out that non-functional testing is difficult. Our objective is to understand to what extent non-functional aspects contribute to quality inflation.
\end{description}

When conducting empirical investigations to address O1 and O2, it became apparent that although traceability links between artifacts (i.e., BR, IR, and test cases) were assumed to exist, this was not necessarily the case.



\subsection{Development Lifecycle at \E}
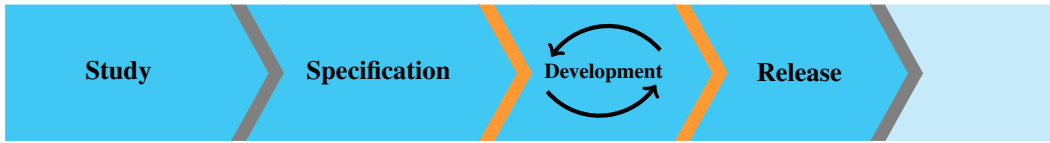
\begin{figure}
    \begin{tikzpicture}
\filldraw[cyan!60] (0,0) rectangle (3,1.8) node[pos=.5, text=black, align=center] {\footnotesize\textbf{Study}};
\filldraw[cyan!60] (3,0) -- (3.5,0.9) -- (3,1.8) -- cycle;
\filldraw[gray] (3,0) -- (3.2,0) -- (3.7,0.9) -- (3.2, 1.8) -- (3,1.8) -- (3.5,0.9);
\filldraw[cyan!60] (3.2,0) -- (3.7,0) -- (3.7,0.9);
\filldraw[cyan!60] (3.2,1.8) -- (3.7,1.8) -- (3.7,0.9);
\filldraw[cyan!60] (3.7,0) rectangle (6.3,1.8) node[pos=.5, text=black, text width=2cm] {\footnotesize\textbf{Specification}};
\filldraw[cyan!60] (6.3,0) -- (6.8,0.9) -- (6.3,1.8) -- cycle;
\filldraw[orange!80] (6.3,0) -- (6.5,0) -- (7,0.9) -- (6.5, 1.8) -- (6.3,1.8) -- (6.8,0.9);
\filldraw[cyan!60] (6.5,0) -- (7,0) -- (7,0.9);
\filldraw[cyan!60] (6.5,1.8) -- (7,1.8) -- (7,0.9);
\filldraw[cyan!60] (7,0) rectangle (8.9,1.8) node[pos=.5, text=black, text width=2cm, align=center] {\scriptsize\textbf{Development}};
\filldraw[cyan!60] (8.9, 0) -- (9.4, 0.9) -- (8.9, 1.8) -- cycle;
\filldraw[orange!80] (8.9,0) -- (9.1, 0) -- (9.6, 0.9) -- (9.1, 1.8) -- (8.9, 1.8) -- (9.4, 0.9);
\draw[black, ultra thick, ->] (8.7,1.2) arc (40:147:0.90);
\draw[black, ultra thick, ->] (7.2,0.65) arc (40:147:-0.90);
\filldraw[cyan!60] (9.1, 0) -- (9.6, 0) -- (9.6, 0.9);
\filldraw[cyan!60] (9.1, 1.8) -- (9.6, 1.8) -- (9.6, 0.9);
\filldraw[cyan!60] (9.6, 0) rectangle (11.5, 1.8) node[pos=.5, text=black, text width=2cm, align=center] {\footnotesize\textbf{Release}};
\filldraw[cyan!60] (11.5, 0) -- (12, 0.9) -- (11.5, 1.8) -- cycle;
\filldraw[gray] (11.5, 0) -- (11.7, 0) -- (12.2, 0.9) -- (11.7, 1.8) -- (11.5, 1.8) -- (12, 0.9);
\filldraw[cyan!20] (11.7, 0) -- (14, 0) -- (14, 1.8) -- (11.7, 1.8) -- (12.2, 0.9);
\end{tikzpicture}
    \caption{Simplified version of the SDLC in use at \E. Orange marks indicate TestOrg main decision points in the process.}
    \label{fig:sdlc}
\end{figure}

A simplified version of \E's development workflow is reported in~\Cref{fig:sdlc}.
The workflow is based on decisions taken together by development and product management and provides a transparent, common understanding of the development status.

\E initiates the \textit{Study} phase to fulfill new market needs, meet their R\&D goals, or respond to a customer request.
By the end of this phase, the stakeholders agree on the scope and resources available for development.
In the \textit{Specification} phase, the development organizations within \E and the organization impacted by the development\footnote{\review{The list of organizations impacted by the development of a new requirement is one of the outcome of the study phase.}}, work on a document specifying a BR, and divide it into Business Sub-Requirements (BSR).
The end of this phase is a synchronization point for the interested organizations, including TestOrg, to align with the BR scope and plan.  
Once the \textit{Development} phase starts, there is an agreement on the development plans, dependencies, and test plans to be carried out.
\review{Inputs to this phase are the BR, the architecture model, general improvements for the area (e.g., mobile payments) and several guidelines (e.g., coding conventions, UX guidelines).}
\textit{Development} ends when TestOrg completes internal verification \review{consisting of unit testing and integration testing.}
\review{TestOrg follows checklists containing the activities necessary to get the BR and BSR to a \textit{done} state.
The outputs of this phase are a package file containing the implemented solution (e.g., a Jar in case of a Java project), the source code, a release note document, and the updated architecture and risk management models.}
This phase is iterative and feedback-driven from internal channels---which perform continuous integration, simulation, and laboratory evaluation---and external ones, including customer laboratory evaluations and restricted launches. 
\review{Each iteration usually lasts  two weeks.}
The end of the \textit{Release} signals that a feature is ready and can be commercially released.

\subsection{Main Development artifacts at \E}
Organizations within \E handle requirements at different levels of granularity, from Business Opportunities (a high-level customer-centric definition of a solution to a problem or need) to User Stories, derived from BSR, implemented by development teams in Scrum sprints.
TestOrg interacts mainly with BRs---i.e., requirements at intermediate granularity---which can be divided into BSR due to size and complexity. 
BR and BSR are specified following a template, and their contents vary in length (on average between 20 and 40 pages).
These specifications include:
\begin{enumerate}[i)]
\itemsep-0.2em 
    \item General Information (e.g., scope, terminology).
    \item Output from the \textit{Study} phase (including recommendation for further studies).
    \item Technical solution description which contains the requirements for the technical use case implementation.
    \item A Glossary of terms used in the document.
    \item References used in the documents.
    \item Changelog and revision information.
\end{enumerate}
The specification documents are stored in a repository that allows tracking of changes. 
A web-based project management tool tracks BR and BSR status, responsible team, and other metadata. 
\review{\Cref{tbl:mingle} shows the main BR attributes tracked using the tool.}
\begin{table}[t]
\caption{BR attributes used in this study extracted from \E's requirements management system.}
\label{tbl:mingle}
\resizebox{\textwidth}{!}{%
\begin{tabular}{@{}lll@{}}
\toprule
\multicolumn{1}{c}{\textbf{Attribute}} & \multicolumn{1}{c}{\textbf{Type}} & \multicolumn{1}{c}{\textbf{Description}}                                                                                                                                               \\ \midrule
name                                   & string                            & name of a requirement                                                                                                                                                                  \\
\rowcolor[HTML]{EFEFEF} 
description                            & text                   & \begin{tabular}[c]{@{}l@{}}description of the requirements. \\ Contains purpose, DoD and references to sub-requirements\end{tabular}                                                   \\
type                                   & emun                              & Indicates the granularity of a requirement—i.e., task, user story, epic                                                                                                                \\
\rowcolor[HTML]{EFEFEF} 
project                                & enum                              & The project the requirement belongs to                                                                                                                                                 \\
version                                & integer                           & Existing revisions of the requirement                                                                                                                                                  \\
\rowcolor[HTML]{EFEFEF} 
created                                & date                              & Date when the requirement is created                                                                                                                                                   \\
modified                               & date                              & Date when the requirement is modified                                                                                                                                                  \\
\rowcolor[HTML]{EFEFEF} 
creator                                & string                            & Name of the employee who created the requirement                                                                                                                                       \\
status                                 & enum                              & Current status of the requirement (released, tested, etc.)                                                                                                                              \\
\rowcolor[HTML]{EFEFEF} 
release                                & enum                              & The targeted release in which the requirement should be included                                                                                                                       \\
business goal                          & enum                              & The targeted high level business requirement                                                                                                                                           \\
\rowcolor[HTML]{EFEFEF} 
validation                             & enum                              & Context in which the requirement is validated                                                                                                                                          \\
planned phase X                        & date                              & \begin{tabular}[c]{@{}l@{}}When the requirement is intended to be moved to the next phase X in the development \\ X $\in$ [study, specification, development, release]\end{tabular} \\
\rowcolor[HTML]{EFEFEF} 
moved to phase X                       & date                              & \begin{tabular}[c]{@{}l@{}}When the requirement is actually moved to the next phase X in the development \\ X $\in$ [study, specification, development, release]\end{tabular}       \\
issue reports                          & list                              & Issue reports currently associated with the requirement \\
\rowcolor[HTML]{EFEFEF}
reference & link & Link to the complete document specification stored in another system) \\
\bottomrule
\end{tabular}
}
\end{table}

Another type of artifact involved in the development process at \E is Issue Report (IR). 
IRs are defects reported before release with varying granularity and can impact several BRs and BSRs.
\E uses a taxonomy of eight software characteristics associated with an IR.
These range from functional suitability (e.g., functional completeness or correctness) to usability and maintainability. 
The IR lifecycle is handled using a web-based issue-tracking tool.
\review{\Cref{tbl:mhweb} show the main IR attributes tracked using the tool.}

Testing activities occur at different levels of abstraction.  
Manual or semi-automated testing is usually performed for BRs that impact several organizations responsible for different DBS solutions within \E.
Manual tests are managed and reported in a separate web-based application.
Automated tests---written in different programming languages and version-controlled in a repository---are also present and implemented to verify (parts of) BRs and BSRs. 
\begin{table}[t]
\centering
\caption{IR attributes used in this study extracted from \E's issue tracker system.}
\label{tbl:mhweb}
\resizebox{\textwidth}{!}{%
\begin{tabular}{@{}lll@{}}
\toprule
\multicolumn{1}{c}{\textbf{Attribute}} & \multicolumn{1}{c}{\textbf{Type}} & \multicolumn{1}{c}{\textbf{Description}}                                 \\ \midrule
name        & string & Name given to the IR                                     \\
\rowcolor[HTML]{EFEFEF} 
content     & string & Content of the IR                                        \\
priority    & enum   & Priority of the IR according to the submitter            \\
\rowcolor[HTML]{EFEFEF} 
registered  & date   & When the IR is created                                   \\
assigned    & date   & When the IR is assigned to a team                        \\
\rowcolor[HTML]{EFEFEF} 
answered    & date   & When a fix for the IR is proposed                        \\
completed   & date   & When the work to address the IR is completed             \\
\rowcolor[HTML]{EFEFEF} 
product     & string & Product experiencing the fault                           \\
market                                 & string                            & Market reference in which the product is experiencing the fault          \\
\rowcolor[HTML]{EFEFEF} 
issuing BR  & string & Name of the BR related to this IR                        \\
characteristic                         & enum                              & Quality characteristic describing the IR (e.g., functional, reliability) \\
\rowcolor[HTML]{EFEFEF} 
hot         & bool   & Whether the IR requires immediate attention              \\
duplicate   & bool   & Whether the IR is a duplicate of an existing one         \\
\rowcolor[HTML]{EFEFEF} 
is child    & bool   & Whether the IR is in a is-a relationship with another IR \\
parent      & string & Reference to the parent IR (iff is child is True)        \\
\rowcolor[HTML]{EFEFEF} 
rejected    & bool   & Whether the IR will be addressed or not                  \\
observation & text   & Free text (e.g., stack traces, steps to reproduce)     \\ 
\bottomrule
\end{tabular}%
}
\end{table}
\section{The Role of Traceability in Investigating Software Quality}
\label{sec:traceability}

\BTH had access to BRs, source code, test results, and IRs. 
Given such a rich set of data, we set out to map artifacts (and their quality) developed early in the Development phase to later ones.
In this section, we present the approach that was initially taken to identify such traces and the challenges that arose during this work.

\subsection{Research Methodology}
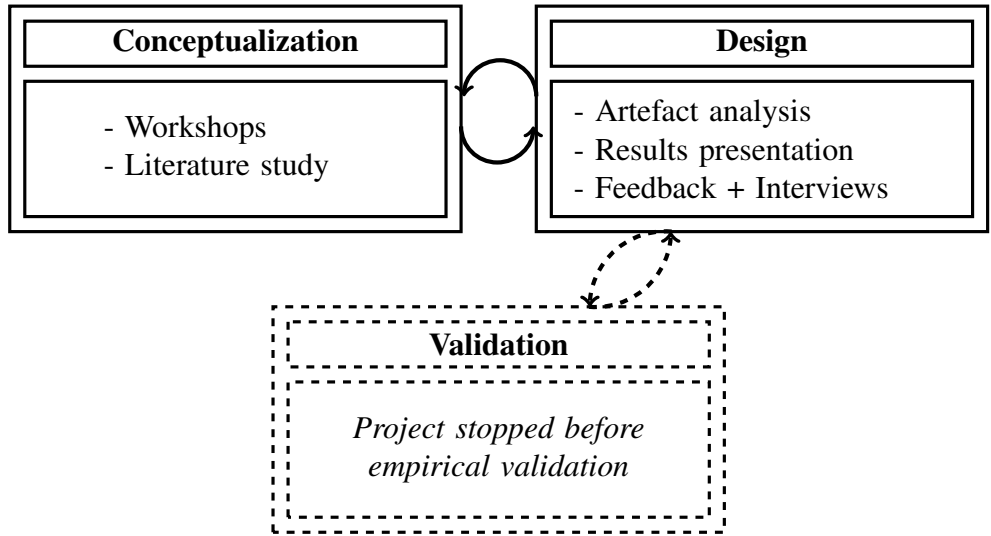
\begin{figure}
    \centering
\begin{tikzpicture}

\draw[black, very thick] (0,0) rectangle (6, 3);
\draw[black, very thick] (0.2,2.2) rectangle (5.8,2.8) node[pos=.5, text=black] {\textbf{Conceptualization}};
\draw[black, very thick] (0.2,0.2) rectangle (5.8,2) node[text width=3.5cm, pos=.5, text=black] {- Workshops\\ - Literature study};

\draw[ultra thick, ->] (7,1.8) arc (0:180:.48);
\draw[ultra thick, ->] (6,1.4) arc (-0:180:-.48);

\draw[black, very thick] (7,0) rectangle (13, 3);
\draw[black, very thick] (7.2,2.2) rectangle (12.8,2.8) node[pos=.5, text=black] {\textbf{Design}};
\draw[black, very thick] (7.2,0.2) rectangle (12.8,2) node[text width=5cm, pos=.5, text=black] {- Artefact analysis\\ - Results presentation \\ - Feedback + Interviews};

\draw[dashed, ultra thick, ->] (8.8,0) arc (85:180:1);
\draw[dashed, ultra thick, ->] (7.7,-1) arc (85:180:-1);

\draw[dashed, black, very thick] (3.5,-4) rectangle (9.5, -1);
\draw[dashed, black, very thick] (3.7,-1.2) rectangle (9.3,-1.8) node[pos=.5, text=black] {\textbf{Validation}};
\draw[dashed, black, very thick] (3.7,-3.8) rectangle (9.3,-2) node[text width=4.7cm, pos=.5, align=center, text=black] {\textit{Project stopped before empirical validation}};
\end{tikzpicture}
    \caption{Overview of design science approach and activities for each phase.}
    \label{fig:design_science}
\end{figure}

During this research collaboration with \E, we followed the design
science paradigm~\cite{RES20} as presented in \Cref{fig:design_science}.
After establishing the research objectives and familiarizing with the existing literature, the researchers at \BTH started the solution design activities.

Initially, the researchers needed to familiarize themselves with the artifacts, domain-specific language, information systems, and processes at \E.
To that end, \BTH obtained access to the infrastructure and worked in collaboration with TestOrg\footnote{One of the researchers spent approximately 20 hours/month at the company office in early 2020 before work-from-home was established due to the pandemic.} to clarify uncertainties and be onboarded on the internal processes and systems in use within the organization.
Moreover, \BTH researchers and TestOrg members had more structured meetings during which the former presented their current understanding of the problem, based on the analysis of the artifact, and proposed alternative ways forward and possible solutions. 
TestOrg gave feedback on the results pointing out, for example, wrong assumptions the researchers made.
The researchers followed up with ad hoc interviews about specific topics when necessary.   
After several iterations between \textit{Conceptualization} and \textit{Design}, \BTH and TestOrg could not produce a solution---i.e., an intervention to reveal where quality inflation was taking place---which could be \textit{Validated} due to the lack of traceability between artifacts. 

\begin{table}[t]
\caption{Metrics used to characterize \textit{Troublesome BR}. In the table, \textit{phase} $\in$ [\textit{study}, \textit{specification}, \textit{implementation}, \textit{release}].}
\footnotesize
\begin{tabular}{ m{4cm} m{9cm} }
\hline
\multicolumn{1}{>{\centering\arraybackslash}m{40mm}}{\textbf{Name}}
&
\multicolumn{1}{>{\centering\arraybackslash}m{70mm}}{\textbf{Definition}}
\\ \hline
Monthly IR for BR & The number of functional IR associated to a specific BR averaged over a monthly period. \\
\rowcolor[HTML]{EFEFEF}  Release IR for BR & The number of functional IR associated to a specific BR averaged over release. \\
Time BR spent in phase & Number of days a BR stayed in a specific phase as indicated by its status. \\
\hline
\end{tabular}
\label{tbl:metrics}
\end{table}

\subsection{Provisional Design Solution}
\label{sec:results}
The initial design goal was a set of guidelines to support TestOrg in identifying issues with quality inflation, inform them about test cases that needed improvement, and suggest BRs that need to be better tested or refined (i.e., reworded, simplified, re-scoped).
Furthermore, we aimed to create automated support through a recommender system, based on such guidelines, which could be integrated into the TestOrg continuous integration environment and reveal ``quality inflated'' BR and BSR.

\begin{figure}[!b]
\centering
\includegraphics[width=\textwidth]{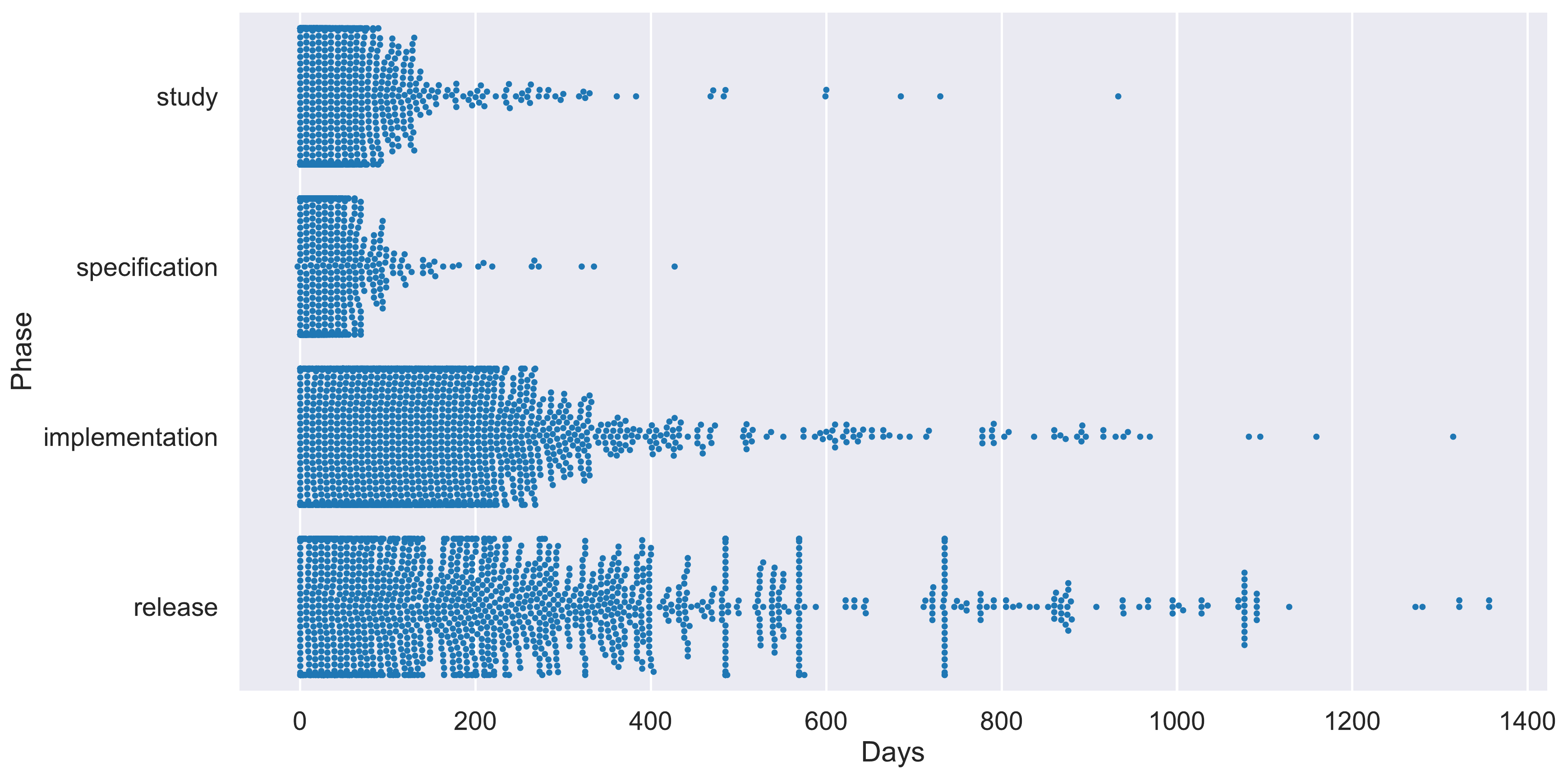}
\caption{BRs times in the different SDLC phases.}
\label{fig:BUC_times}
\end{figure}

Based on the initial conceptualization and discussion with TestOrg, we hypothesize that writing automated tests can be more challenging for some BRs than for others (Objective O1).
In the first design iteration (left pane of~\Cref{fig:tracebility}), we characterize such \textit{Troublesome BRs} using several metrics mined from the \E artifact repositories (\review{see \Cref{tbl:metrics}}).
In particular, we considered the number of IRs associated with a BR over time (e.g., in a release), the amount of time a BR spent in the different phases of the development~(e.g., as reported in \Cref{fig:BUC_times}), and the difference between the planned and actual time for a BR to advance to the next phase. 
When defining the \textit{Troublesome BR} metric, we used the existing traceability link between IR and BR available in the issue tracking tool (marked with \textcircled{1} in \Cref{fig:tracebility}).

In informal interviews, the researchers presented and discussed their results together with TestOrg and got early feedback that helped better conceptualize the solution design.
For the second iteration (right pane of~\Cref{fig:tracebility}), we needed to define the quality of the test cases for a BR to obtain a metric (i.e., \textit{Test suite quality}) which we could then correlate with \textit{Troublesome BR}.
We discussed several ways of establishing the necessary traceability links between BRs and test cases.
The more immediate one---a direct bidirectional link between BR and test cases is unavailable (marked with \textcircled{2} in \Cref{fig:tracebility}).
In their development process, \E does not enforce traceability between high-level requirements, such as BR, and low-level code artifacts, such as test cases. 

\begin{figure}
\centering
    \begin{tikzpicture}

\draw[line width=.2mm, black] (-.5, -.8) rectangle (7, 4.5); 
\node[anchor=west,baseline,inner sep=0.5pt] (a) at (0, -0.5) {\textit{1st iteration}};

\draw[line width=.2mm, black] (7.5, -.8) rectangle (14.2, 4.5); 
\node[anchor=west,baseline,inner sep=0.5pt] (a) at (7.9, -.5) {\textit{2nd iteration}};

\draw[black, ultra thick] (0,0) -- (2.3,0) -- (2.3,1.1) -- (2, 1.4) -- (0, 1.4) -- cycle; 
\node[anchor=east,baseline,inner sep=0.5pt] (a) at (1.5, 0.8) {\textbf{IR}};

\draw[black, ultra thick] (0,2.5) -- (2.3,2.5) -- (2.3, 3.6) -- (2, 3.9) -- (0, 3.9) -- cycle;

\node[anchor=east,baseline,inner sep=0.5pt] (c) at (1.6, 3.3) {\textbf{BR}};
\node[anchor=east,baseline,inner sep=0.5pt] (d) at (1.5, 2.9) {\tiny History};

\node[anchor=east,baseline,inner sep=0.5pt] (c) at (6.6, 2.4) {\small\textbf{\textit{Troublesome}}};
\node[anchor=east,baseline,inner sep=0.5pt] (d) at (5.9, 1.9) {\textbf{\textit{BR}}};

\filldraw[fill=red!0, very thick](1.4,2) circle (.2);
\node[anchor=east,baseline,inner sep=0.5pt] (b) at (1.54, 2.03) {\textbf{1}};

\draw[line width=1mm, gray] (4, 1.5) rectangle (6.8, 2.9); 
\draw[ultra thick, gray, ->] (2.3,0.7) arc (60:155:-1.6);
\draw[ultra thick, gray, ->] (2.3,3.4) arc (120:39:1.8);
\draw[black, ultra thick, ->] (1.1, 1.4) -- (1.1, 2.5);

\draw[line width=1mm, gray] (8, 1.5) rectangle (10.2, 2.9);
\node[anchor=east,baseline,inner sep=0.5pt] (a) at (9.5, 2.6) {\textbf{\textit{Test}}};
\node[anchor=east,baseline,inner sep=0.5pt] (a) at (9.6, 2.2) {\textbf{\textit{suite}}};
\node[anchor=east,baseline,inner sep=0.5pt] (b) at (9.9, 1.8) {\textbf{\textit{quality}}};

\draw[ultra thick, black] (11.5,-.5) -- (13.7,-.5) -- (13.7, .6) -- (13.4, .9) -- (11.5, .9) -- cycle;
\node[anchor=east,baseline,inner sep=0.5pt] (a) at (13.61, 0.4) {\textbf{IR-fixing}};
\node[anchor=east,baseline,inner sep=0.5pt] (b) at (13.42, -0.1) {\textbf{commit}};
\draw[ultra thick, black, ->] (12.5, 0.9) -- (12.5, 1.5);

\draw[ultra thick, black] (11.5,1.5) -- (13.7,1.5) -- (13.7, 2.6) -- (13.4, 2.9) -- (11.5, 2.9) -- cycle;
\node[anchor=east,baseline,inner sep=0.5pt] (a) at (13, 2.4) {\textbf{Test}};
\node[anchor=east,baseline,inner sep=0.5pt] (b) at (13.1, 2) {\textbf{cases}};
\filldraw[fill=red!0, very thick](12,3.5) circle (.2);
\node[anchor=east,baseline,inner sep=0.5pt] (b) at (12.15, 3.52) {\textbf{2}};

\filldraw[fill=red!0, very thick](10.3,.6) circle (.2);
\node[anchor=east,baseline,inner sep=0.5pt] (b) at (10.45, .63) {\textbf{3}};
\filldraw[fill=red!0, very thick](12.9,1.2) circle (.2);
\node[anchor=east,baseline,inner sep=0.5pt] (b) at (13.05, 1.2) {\textbf{4}};

\draw[ultra thick, gray, ->] (11.5,2.2) -- (10.28, 2.2);

\draw[black,ultra thick, ->, dashed] (8,2.4) -- (6.9, 2.4);
\draw[black, ultra thick, ->, dashed] (6.9, 1.9) -- (7.9, 1.9);

\draw[black, ultra thick, <->, dashed] (2.3, .3) -- (11.5,.3);
\draw[black, ultra thick, <->, dashed] (12.6, 2.9) arc (64:107.3:14.2);

\draw[ultra thick, gray] (1.6, 5) -- (3, 5);
\node[anchor=west] (l) at (0, 5) {\textit{Metrics}};

\draw[ultra thick, black] (5.8, 5) -- (7.2, 5);
\node[anchor=center] (l) at (4.7, 5) {\textit{Traceability}};

\draw[ultra thick, black, dashed] (10.6, 5) -- (12, 5);
\node[anchor=center] (l) at (9.2, 5) {\textit{No traceability}};

\end{tikzpicture}
    \caption{Traceability (and lack thereof) between artefacts used for the proposed design.}
    \label{fig:tracebility}
\end{figure}

Next, since we already had a traceability link between IR and BR, we investigated downstream (i.e., IR-to-source code) and upstream (i.e., source code-to-IR) traceability links to connect IR and test cases via IR-fixing commits (marked with \textcircled{3} in \Cref{fig:tracebility}).
From a commit, it is possible to establish a link to the test cases (marked with \textcircled{4} in \Cref{fig:tracebility})---e.g., using the approach proposed in \cite{ZVD08}.
However, the field that explicitly connects IR and commit in the issue-tracking system is barely used (approximately 10\% of IRs contains a reference to a commit). 

We then looked at the upstream traceability between test cases and IRs, using information from the IR-fixing commit (the other direction of the arrow marked as \textcircled{3} in \Cref{fig:tracebility}). 
Development organizations within \E ~\review{are required to use a} structured commit template.
The template has a field in which the developer can indicate, among other things, whether the commit is part of a fix. The developer can do this by including the \textit{id} of the artifact describing the issue, such as an IR.
From the source code version control system, we mined patches and associated discussions taking place during the same timeframe during which BRs were implemented and IRs were addressed (i.e., 2016--mid 2020).
We parsed the commit messages looking for fixes mentioning IRs and their id. 
However, this automated approach returned a low number of hits.
Upon manual inspection of 50 random commit messages, it appeared that the commit template is mostly filled in by automatic tools (e.g., static code analyzers) or used for tracing refactoring to code smells (e.g., from SonarQube) which are outside the scope of this study.


\review{In summary, \Cref{tbl:gqm} show how the goal of the study relates to the metrics using the GQM framework~\cite{BCR94}.
In particular, we defined troublesome BRs but could not establish the QA effort associated with them due to lack of traceability.  
In turn, we could not answer the remaining questions to fulfill our original goal.}

\begin{table}[t]
\caption{Goal-Question-Metrics for the study of quality inflation at \E. }
\tiny
\begin{tabular}{m{3cm} m{4cm} m{3cm} m{2cm}}\\
\midrule
\multicolumn{1}{>{\centering\arraybackslash}m{30mm}}{\textbf{Goal} (based on \textit{O1} and \textit{O2})}
&
\multicolumn{1}{>{\centering\arraybackslash}m{40mm}}{\textbf{Question}}
& 
\multicolumn{1}{>{\centering\arraybackslash}m{30mm}}{\textbf{Metrics}}
& 
\multicolumn{1}{>{\centering\arraybackslash}m{20mm}}{\textbf{Answered?}}
\\
\midrule
\multirow{4}{1cm}[-2.5em]{
    \begin{tabular}[c]{@{}l@{}}
        \textit{Purpose}: Characterize the\\ 
        \textit{Issue}: spread of\\ 
        \textit{Object}: quality inflation\\ 
        \textit{Viewpoint}: QA managers at\\ 
        \textit{Context}: \E
    \end{tabular}} 
& How widespread is quality inflation? & $\frac{\text{Quality inflated TR}}{\text{All TR}}$ & No \\
& When is there  a mismatch between QA effort and quality observed for a BR?  & $\frac{\text{QA effort for TR}}{\text{Observed Quality for TR}} > ~\text{Threshold} $ &No \\
& What is the QA effort dedicated to a BR? & 
\begin{tabular}[c]{@{}l@{}} Test smells\\  Test suite effectiveness \\ Test suite time to complete\\  Defect density\\  Defect age \end{tabular}
& \textbf{No, due to lack of traceability}\\  &  Why are some BR perceived to be more troublesome than others? & 
\begin{tabular}[c]{@{}l@{}} Monthly IR for BR\\  Release IR for BR\\  Time BR spent in phase 
\end{tabular} 
& Yes
\\
\bottomrule
\end{tabular}
\label{tbl:gqm}
\end{table}
\section{Discussion}
\label{sec:discussion}
\begin{figure}
\begin{centering}
\begin{tikzpicture}[scale=.98]
  \path[mindmap,concept color=blue!60,text=white, font=\sf, minimum size=0pt, text width=2.2cm]
    node[concept] {Lack of Traceability}
    child[concept color=green!50!black] {
      node at (50:6)[concept, minimum size=0pt, text width=2.2cm] {Consequences}
      [clockwise from=120]
      child { node[concept] {Integration issues} }
      child { node[concept] {Lack of ownership} }
      child { node[concept] {Redundant work} }
      child { node[concept, minimum size=0pt, text width=1.7cm] {Complex defect\\ management } }
      child { node[concept] {Inefficient resource allocation} }
    }  
    child[concept color=red]  {
      node at (-50:-6)[concept, text width=1.5cm, minimum size=0pt] {Causes}
      [clockwise from=-54]
      child { node[concept] {Wrong feature scoping} }
      child { node[concept] {Outdated legacy system} }
      child[concept color=red!30!brown] {
      node[concept, text width=1cm, minimum size=0pt]{Tools}
      [counterclockwise from=110]
        child {node(a)[concept, minimum size=0pt, text width=1.3cm] {Interoperability}}
        child {node[concept, below left=2pt and 2pt of a] {Fragmented information}}
      }
      child[concept color=red!70!black] {
      node[concept, minimum size=0pt, text width=1cm] {Human}
      [counterclockwise from=45]
        child{node[concept]{Perceived value}} 
        child{node[concept]{Team norm }} 
        child{node(v)[concept]{Lack of resource}} 
        child{node[concept, below left=2pt and 2pt of v]{Integration of different teams}} 
      }
      child { node[concept, text width=1.8cm] {Components\\ integration} }
    };
\end{tikzpicture}
\caption{Causes and consequences of lack of traceability.}
\label{fig:mindmap}
\end{centering}
\end{figure}
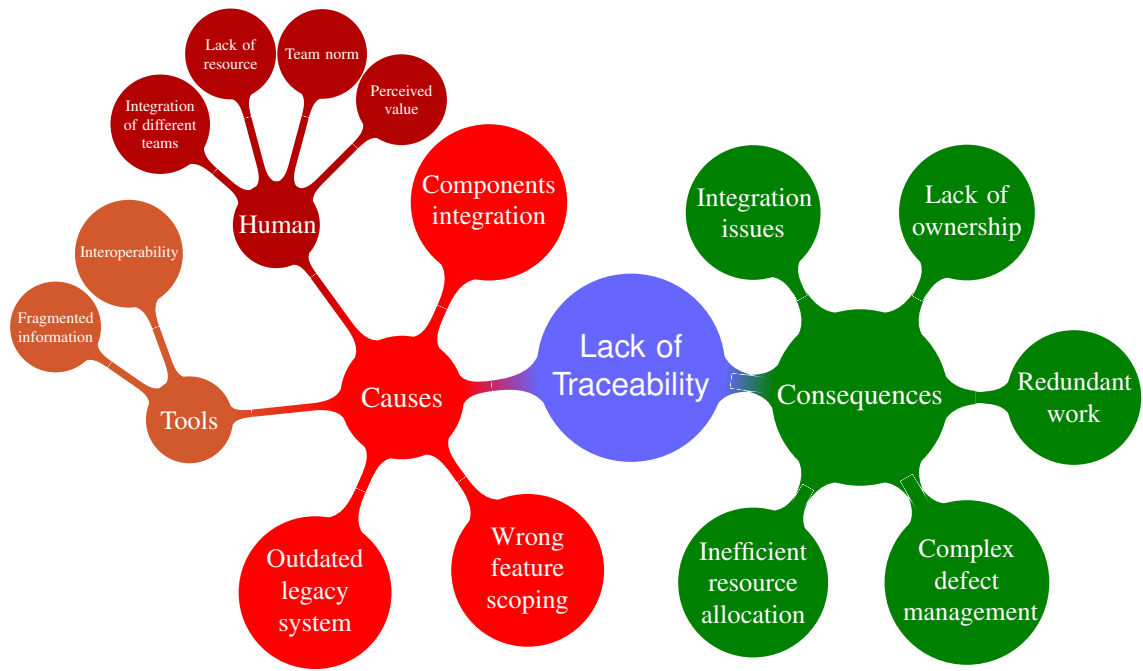

In this section, we provide the lessons learned from our failed attempt at studying (and eventually addressing) quality inflation at \E~\review{due to lack of traceability} and discuss the implications of these results for research in this field. 
\review{We summarize our considerations about the causes and consequences of lack of traceability in~\Cref{fig:mindmap}.}

\review{The lessons learned and considerations presented in this section are derived from our own experience analyzing the artifacts used within \E and our interactions with practitioners in TestOrg during workshops, feedback, and interview sessions. 
The takeaways are complemented with references to the literature that can provide further insights.}
\subsection{Lessons learned}
We report the lessons learned from practitioner and researcher perspectives.
For the latter, we include takeaways that should be considered when studying traceability in complex industry settings.

\paragraph{For practitioners}
In the context of a company, the traceability links are maintained for a purpose that can be different from the one of a research project.
We recommend that establishing and updating traceability links, at least the one that matters for the company, should be treated as a backlog item and tracked like any other items in the development process. 
Moreover, the explicit lack of traceability needs to be treated as a technical debt item and included in  Sprints aimed at paying it back.

\review{Toledo et al~\cite{TMS21} show that the lack of traceability between artifacts leads to architectural technical debt.
They show how, in the case of microservice architectures, maintaining data-to-data source traceability is often necessary to fulfill regulations and identify services that are not needed anymore.
Whereas, also in the area of architecture and design, Charalampidou et al~\cite{CAC18} show that it is useful to document traceability links to manage and estimate the cost of paying debt.
Similarly, traceability supports managing documentation and requirements debt~\cite{CAE21}; conversely, lack thereof is detrimental to maintenance tasks.
}

Considering traceability is important as lack of links, or hard-to-establish links, can be \textit{smells} for other problems, such as wrong scoping. 
For example, when a requirement is too large in scope, its implementation is expected to receive several change requests. However, understanding the scope requires a mechanism to trace change requests to requirements.

Traceability of artifacts is important from a management perspective. 
Without it, several overhead costs can be expected due to lack of implementation ownership---i.e., which team is responsible for  implementing a functionality \review{(for example, see~\cite{DBA13, ABB20})}. 
This lack of ownership can, in the worst case, lead to the same functionality being implemented several times or for the implementation to be disrupted during integration due to lacking knowledge of the code dependencies~\review{\cite{DAM19}}.

Traceability also helps mitigate failure propagation---i.e., due to defects in the code that endure through the development cycle and potentially reach the customer \review{(for example, see~\cite{ME15,CBS16}).}
Aligning tests with requirements to establish coverage metrics is vital, and without this information, it can be unclear if a requirement has been correctly tested or not. 
Despite defects reaching the customer or not, lingering faults cause additional overhead, delay releases, and result in longer implementation time.

Traceability also helps mitigate uncertainties regarding the allocation of resources \review{(e.g., ~\cite{WPS21})}.
While changed or added requirements give input to allocate more resources, verification of said requirements provides grounds for their deallocation. 
However, without knowing if a development task has been properly addressed, such deallocation can be delayed or spent inefficiently \review{(for example, see~\cite{Alp19,AE20,STD19}).}

\paragraph{For researchers}
Researchers should validate their assumptions about what is available in terms of traceability when collaborating with a large company.
Some activities that are taken for granted in some settings (e.g., open source) are hard to apply in a complex industrial organizational context, such as \E. 
In such contexts, it is inherently difficult to have an overview of what is available in terms of information, data, and artifacts, and what is not. 
In the case of TestOrg, although the managers were aware of the traceability between BRs and IRs, the lack of traceability links between the source code and IRs or BRs was not considered since it is outside the scope of their activities. 
For the researchers, this became clear in discussions with people in more operational roles.

\takeaway{
When assessing traceability links between different artifacts involve early personnel working day-to-day with such artifacts. For example, when traceability between source code and requirements is needed,  developers and business analysts in the company should be involved.
}
Within the organization, we realised that horizontal traceability (i.e., traceability between artifacts at the same level of abstraction, such as requirement-to-requirement) has more value than vertical traceability (e.g., at different levels, such as test cases-to-requirements). 
This may be the case in large companies, where the different phases of the development cycle---and, therefore, their associated artifacts---are managed by different internal organizations.
For the practitioners we interacted with, limited traceability (mostly among BRs, and between BRs and IRs) was enough to perform their daily tasks. 
To enact a different type of traceability, researchers need a strong use case for the company. 
The company will have to i) allocate resources to support the researchers in establishing  extra traceability links, ii) maintain the traceability links (e.g., for further evaluation by the researchers).

\takeaway{We recommend gaining an early understanding of the organizational structure, and being aware that the amount of traceability information available may be influenced by such structure.}
\review{\textit{Take-away 1} and \textit{Take-away 2} are related to research on Conway's law (e.g.,~\cite{HG99a,HG99b}) and the impact that traceability has on organizational vs. architectural structure. 
Moreover, research on \textit{communication} of traceability highlights how organizational structure influences the way tasks are allocated and \textit{who} within the company possesses the necessary knowledge about artifacts of interest for the researcher~\cite{imtiaz2011effective}.}

In the context of a large organization, different artifacts are tracked at different levels of detail.
In our collaboration with \E, we realized that the system used to track BRs was not populated with much information (e.g., many fields were left blank or filled with boilerplate values).
The organizational \textit{norm} in case BR details are needed is to refer to the BR specification document (through a link in the tracking system) stored in a separate repository.
This limited the automatic extraction of information from BRs, as the two systems are not designed to communicate autonomously. 

\takeaway{
When establishing traceability, consider that in complex settings different information about the same artifact are likely to be scattered across systems. 
Regardless of the approach, consolidating such information requires knowledge of the company's norms.}
\review{
Several researchers tried to address the challenge of scattered information~\cite{SWH21,WSH21,TMP13}.
Promising approaches have combined traditional information retrieval with model-driven engineering~\cite{sannier2012toward} and (semi)supervised machine learning~\cite{BCG19}.}

TestOrg was aware of the problem with dispersed information.
They led, within \E, an initiative to centralize test cases and BR tracking for several purposes, including improving their traceability. 
By the end of such initiative, \E will use a single tracking system for all these artifacts.

\takeaway{Depending on the structure and location of the information required to establish traceability, a fully-automated solution may not be feasible. A solution for creating traceability links should not start by considering full-fledged automation but by accommodating human intervention.}
Tool support is fundamental when establishing, using, and maintaining traceability links. 
Organizations developing complex systems deal with artifacts at different levels of abstraction, details, and formats, which entail using different artifact-tracking tools.
Moreover, practitioners choose, configure, and use tools according to the level of traceability necessary for \textit{their} tasks---for them interoperability may not be a decisive criterion for selecting a tool. 
For example, \E uses an homebrew system for tracking requirements specification documents and IRs, a third-party commercial solution (which reached end-of-life in early 2019, but it is still maintained for legacy reasons) for tracking BRs, and different open-source systems for source code version control and code reviews. 
Some offer APIs, but none offered out-of-the-box integration towards any of the other systems. 

\takeaway{
Fragmentation in terms of tools within a company developing complex systems is to be expected. 
Therefore, effort is required to achieve interoperability between systems when establishing traceability links.}
\review{Addressing tools fragmentation is significant for safety-critical software development~\cite{baumgart2014recipe} and could be mitigated, for instance, through modelling~\cite{drivalos2008engineering}.}

In the case of \E, fragmentation (and the consequent lack of traceability) derived from a tradeoff between other system properties deemed more desirable. 
During our feedback sessions with TestOrg, it became apparent that the homebrew solution was selected to have control over i) the security of the specifications as they contain information pivotal for \E's competitive advantage, and ii) the redundancy of the storage to avoid costly data loss.

\subsection{Considerations on traceability in complex industrial settings}
The results of this study provide insights into the state-of-practice, lessons learned, and considerations for industrial practitioners and academics to reflect upon regarding traceability.
The experience reported in this paper highlights a dichotomous and puzzling situation that was surprising for the researchers involved in this study.
The academic literature on traceability and alignment supports the idea that  traceability among software artifacts is needed to manage complexity~\cite{FWK17}.
Despite these claims, \E is producing systems in the order of millions of lines of code, with thousands of developers, yet traceability links from high-level requirements to source code and tests are not readily available.
Paradoxically, traceability within \E may not be achieved because of the system's complexity, while traceability could mitigate said complexity.

What is going on in this case?
The results may seem baffling, but in real-world settings---when software grows and the organization along with it---several factors influence the evolution of a product, its development process, and the environment.
We did not study the root cause of the current situation, but several hypotheses can be formulated.

First, the system is considered a system of systems (SoS), developed in a heterogeneous environment of processes, tools, and third-party components.
Without a strong culture to tie this development together with an emphasis on traceability, it is only natural that its amount and consistency will vary.
Integration and SoS development are still considered possible, as such is managed at a higher level of abstraction, primarily considering the interfaces of the underlying components, despite the lack of end-to-end traceability.


Second, achieving traceability in a large system is resource-intensive, and human commitment also comes into play.
In a study by Borg et al. about change requests and change management, it was observed that teams are hesitant to even touch upon other teams' code~\cite{borg2017software}.
In larger silo organizations this situation is exacerbated since silos may assume that other individuals or teams keep traceability up to date, especially when teams suffer from resource constraints.

Take, as an example, a scenario in which Team A is adapting a system core component to conform to a BR.
As BRs are high level, they likely require changes to surrounding components.
The developers in Team A make changes, but due to the lack of insight or knowledge about the Team B artifacts, not all artifacts associated with the modified components are updated.
This causes a slight misalignment between the requirement traces and the system under development.
After many such changes, traceability will naturally degrade over time, leading to a situation where traces will have to be maintained or reverse engineered.
However, since reverse engineering is expensive, process solutions that enforce localized knowledge---i.e., silo organizations are instead encouraged.

Finally, there can be a legacy component to the challenge---i.e., that components within the SoS are decaying.
These components may have had extensive traceability information, but as Agile practices are leaner in terms of documentation, such traces are no longer maintained.
Hence, a situation that is perceived as a product of a new way of working with software in which artifacts of long-term value---associated with high cost---are de-prioritised in favour of short-term value gains and other forms of light-weight documentation that fulfill similar purposes.


Furthermore, we observed that traceability between high-level requirements, source code, and test suites is not maintained at \E.
A question remains, \textit{how can \E continue producing high-quality, large-scale, and complex software on time while keeping their customers satisfied?}
As discussed in this section, we believe it is due to the evolution of the development organization and its ability to adapt to circumstances in which traceability is not always available.
Hence, instead of relying on traceability information, workarounds and alternate processes, coupled with organizational structures and architectural design decisions, provide \E a cohesive understanding of the system.

Hence, although the academic literature highlights the need for traceability for understanding of how the system fits together~\cite{kukkanen2009applying}, the situation we observed at \E indicates that such requirements may have been overstated.



\section{Conclusion}
\label{sec:conclusion}
In this paper, we reported our experience in applied research with \E. 
In particular, our goal was to define and apply an intervention for identifying quality inflation using artifacts in different development phases.
However, our attempt failed due to the lack of traceability between such artifacts.

We reflected on the outcomes of this industrial study, which led us to question the  role and perceived value of traceability in industry, and its return-on-investment for large software projects. 
\review{From our experience, we encourage practitioners to be selective about the traceability links necessary for their organization while highlighting the importance of requirement-to-test case trace links.
Moreover, practitioners need to explicitly manage traceability links as they do for other artifacts (e.g., requirements, documents, test cases).} 
We also suggest takeaways that can support researchers performing empirical studies that consider traceability a prerequisite. 
\review{In particular, they need to be aware that the organizational structure and norms can impact which traceability links exist and how they are managed.
For complex projects in large settings, scattered information among different tools and systems is to be expected.}

In the future, we aim to systematically study the state-of-practice related to traceability among our industrial partners. 

\section{Acknowledgements}
Davide Fucci and Emil Alégroth would like to acknowledge that this work was supported by the KKS foundation through the S.E.R.T. Research Profile project at Blekinge Institute of Technology.

\bibliographystyle{elsarticle-num} 
\bibliography{bibliography}





\end{document}